\documentclass[lettersize,journal]{IEEEtran}
\usepackage{booktabs}
\usepackage{amsmath,amsfonts}
\usepackage{array}
\usepackage[caption=false,font=scriptsize,labelfont=rm,textfont=rm]{subfig}
\usepackage{textcomp}
\usepackage{stfloats}
\usepackage{url}
\usepackage{verbatim}
\usepackage{graphicx}
\usepackage{cite}
\usepackage{CJKutf8}
\usepackage{amsthm,amsmath,amssymb} 
\usepackage{algpseudocode}   
\usepackage{mathrsfs}
\usepackage{multicol}
\usepackage{svg}
\usepackage{graphicx}
\usepackage[ruled,linesnumbered]{algorithm2e}
\usepackage{algpseudocode}
\usepackage[colorlinks,linkcolor=blue,anchorcolor=blue,citecolor=blue]{hyperref}

\begin{document}

\title{Dual-Polarization Stacked Intelligent Metasurfaces for Holographic MIMO}

\author{Yida Zhang, Qiuyan Liu, Hongtao Luo, Yuqi Xia, Qiang Wang
\thanks{This paper was partially funded by the National Natural Science Foundation of China (No.62071066 and No.62327801), Fundamental Research Funds for the Central Universities (2242022k60006) and the National Key R \& D Program of China(2020YFB1806602). (Corresponding author: Qiang Wang.)}
\thanks{Yida Zhang, Hongtao Luo, Yuqi Xia, Qiang Wang are with the National Engineering Research Center for Mobile Network Technologies, Beijing University of Posts and Telecommunications, Beijing 100876, China (e-mail: \{zhangyida02, mashirokaze1971, xiayuqi, wangq\}@bupt.edu.cn)}
\thanks{Qiuyan Liu is with the China United Network Communications Corporation Research Institute, Beijing 100037, China (e-mail: liuqy95@chinaunicom.cn)}
\thanks{Simulation code and picture: \url{https://github.com/ZhangYida02/Dual_Polarization_Stacked_Intelligent_Metasurfaces_for_Holographic_MIMO.git}
}}



\maketitle

\begin{abstract}
To address the limited wave domain signal processing capabilities of traditional single-polarized stacked intelligent metasurfaces (SIMs) in holographic multiple-input multiple-output (HMIMO) systems, which stems from limited integration space, this paper proposes a dual-polarized SIM (DPSIM) architecture. By stacking dual-polarized reconfigurable intelligent surfaces (DPRIS), DPSIM can independently process signals of two orthogonal polarizations in the wave domain, thereby effectively suppressing polarization cross-interference (PCI) and inter-stream interference (ISI). We introduce a layer-by-layer gradient descent with water-filling (LGD-WF) algorithm to enhance end-to-end performance. Simulation results show that, under the same number of metasurface layers and unit size, the DPSIM-aided HMIMO system can support more simultaneous data streams for ISI-free parallel transmission compared to traditional SIM-aided systems. Furthermore, under different polarization imperfection conditions, both the spectral efficiency (SE) and energy efficiency (EE) of the DPSIM-aided HMIMO system are significantly improved, approaching the theoretical upper bound.
\end{abstract}

\begin{IEEEkeywords}
Dual-polarization, SIM, DPSIM, HMIMO, RIS, Wave based computing.
\end{IEEEkeywords}

\section{Introduction} 
Over the past decades, multiple input multiple output (MIMO) antenna architectures have evolved significantly to meet growing data throughput and connection density demands. Conventional fully-digital antenna arrays mitigate inter-antenna interference (IAI) through baseband digital precoding and combining, ensuring independent transmission of data streams~\cite{4549739}. However, this approach requires many radio frequency (RF) chains, leading to excessive energy consumption and prohibitive hardware complexity~\cite{5783993}. To address these challenges, a novel antenna that integrates reconfigurable intelligent metasurfaces (RISs) in the transceiver has been proposed. By precisely controlling the electromagnetic (EM) units in the RIS, the transceiver can perform joint signal processing in the digital domain and wave domain. This architecture is called Holographic MIMO (HMIMO), which minimizes the number of RF chains and only requires the use of low-resolution digital-to-analog converters (DACs) and analog-to-digital converters (ADCs), significantly reducing the hardware cost and improving energy efficiency (EE)~\cite{10232975}.

In recent years, most existing HMIMO research is focused on single-layer metasurface structures~\cite{10232975}. To further improve wave domain signal processing capability, ~\cite{10158690} first proposed the HMIMO system assisted by stacked intelligent metasurfaces (SIM), significantly enhancing system performance compared to traditional massive MIMO systems and RIS-assisted HMIMO systems. Based on this,~\cite{li2025stackedintelligentmetasurfacesenhancedmimo} explores the potential of SIM in wideband HMIMO systems, achieving IAI-free multicarrier transmission under frequency-selective fading.~\cite{10922857} studies the application of SIM in multi-user beamforming, achieving a higher spectral efficiency (SE) and lower energy consumption than traditional MIMO systems. Furthermore, SIM has been combined with areas such as integrated sensing and communication (ISAC)~\cite{10803090} and semantic communication~\cite{huang2024stackedintelligentmetasurfacestaskoriented}. However, restricted by the limited integration space at the transceiver, the number of SIM metasurface layers and the scale of EM units cannot be arbitrarily expanded. Consequently, further improving HMIMO systems' performance under the same spatial scale remains an open problem.


Recently, a novel dual-polarized RIS (DPRIS) has been successfully realized~\cite{Ke2021Linear},~\cite{Liu2016Anisotropic},~\cite{Zhang2020Polarization}, where each dual-polarization electromagnetic (DPEM) unit can independently control the phase of signals with different polarizations and exhibits high isolation.~\cite{9497725} points out that DPRIS can effectively suppress polarization cross-interference (PCI) and MIMO inter-stream interference (ISI).~\cite{10675359} investigated the performance gain brought by DPRIS-assisted HMIMO systems compared to RIS-assisted systems. Inspired by this, the main contributions of this paper are as follows:
\begin{enumerate}[]
    \item To further improve HMIMO system performance within the same spatial constraints, we propose the design of a dual-polarized SIM (DPSIM) to enable simultaneous wave-domain processing in two orthogonal polarization directions. Subsequently, we provide a mathematical model for the DPSIM-assisted HMIMO system.
    \item To achieve end-to-end data stream spatial multiplexing, we propose the layer-by-layer gradient descent with water-filling (LGD-WF) algorithm to actively construct matched channels in DPSIM. Consequently, multiple data streams can be transmitted and received directly without precoding processing.
    \item The performance of DPSIM-assisted HMIMO systems is evaluated and compared with that of SIM-assisted systems and conventional massive MIMO systems in terms of ISI, SE, and EE.
\end{enumerate}



\section{System Model}

\begin{figure}[!t]
\centering
\setlength{\abovecaptionskip}{-0.35cm}
\includegraphics[width=3in]{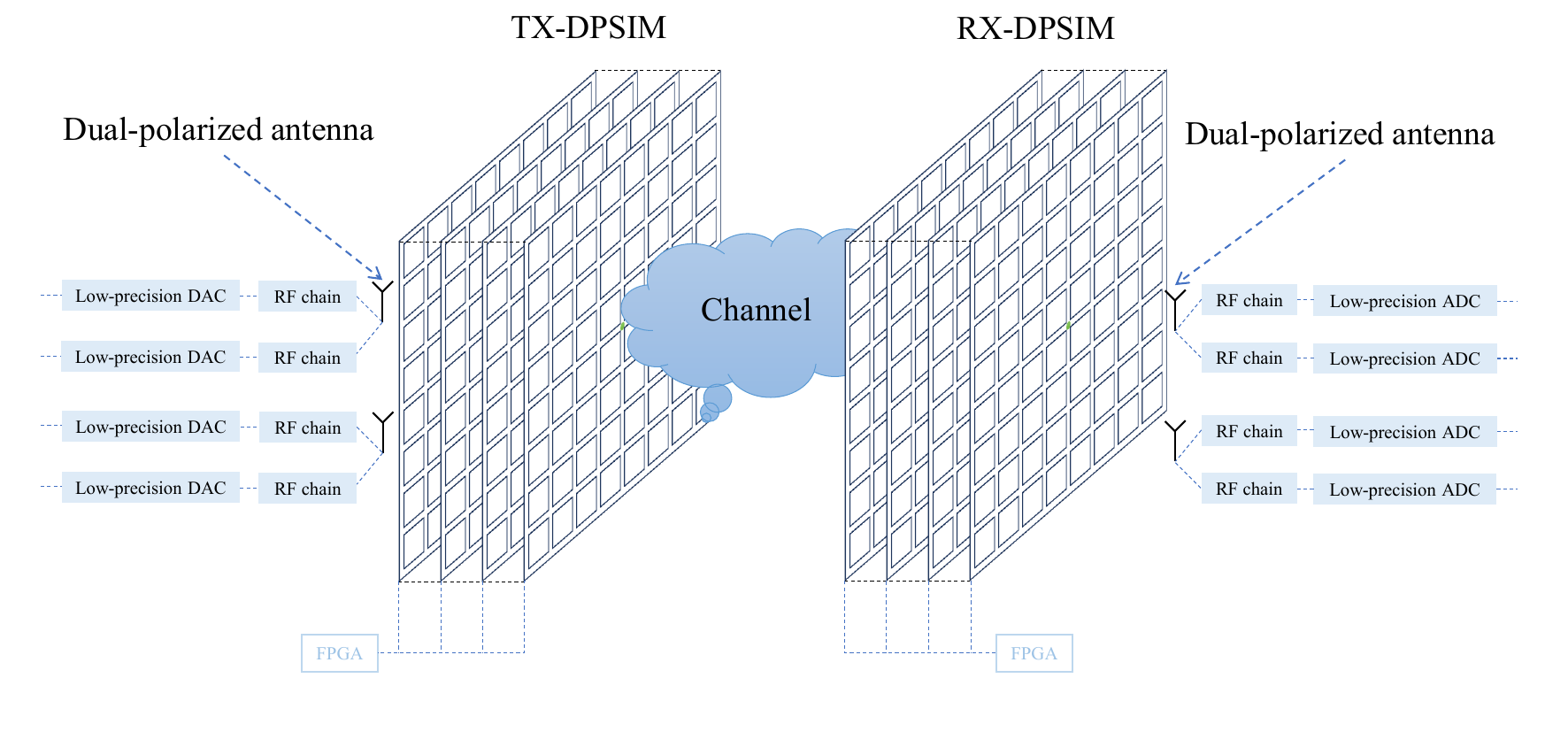}
\caption{Schematic diagram of the DPSIM-assisted HMIMO system.}
\label{Fig1}
\vspace{-0.45cm}
\end{figure}

\begin{figure}[!t]
\centering
\setlength{\abovecaptionskip}{-0.4cm}
\includegraphics[width=3.1in]{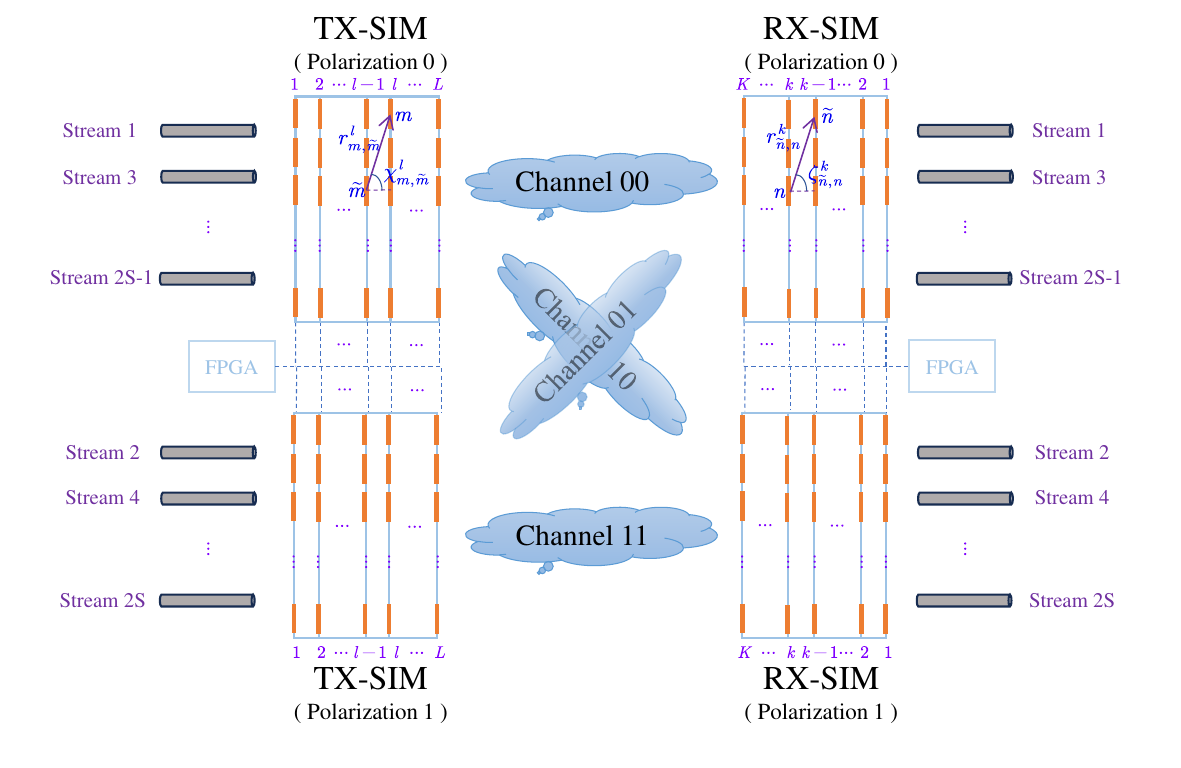}
\caption{The schematic diagram of the DPSIM-assisted HMIMO system in the two orthogonal polarization directions.}
\label{Fig2}
\vspace{-0.1cm}
\end{figure}

\begin{figure*}[!b]
\begin{small}
\vspace{-1cm}
\begin{align*}
&\overline{\ \ \ \ \ \ \ \ \ \ \ \ \ \ \ \ \ \ \ \ \ \ \ \ \ \ \ \ \ \ \ \ \ \ \ \ \ \ \ \ \ \ \ \ \ \ \ \ \ \ \ \ \ \ \ \ \ \ \ \ \ \ \ \ \ \ \ \ \ \ \ \ \ \ \ \ \ \ \ \ \ \ \ \ \ \ \ \ \ \ \ \ \ \ \ \ \ \ \ \ \ \ \ \ \ \ \ \ \ \ \ \ \ \ \ \ \ \ \ \ \ \ \ \ \ \ \ \ \ \ \ \ \ \ \ \ \ \ \ \ \ \ \ \ \ \ \ \ \ \ \ \ \ \ \ \ \ \ \ \ \ \ \ \ }\\
\label{eq1}
&\ \ \ \ \ \ \ \ \ \ \ \ \ \ \ \ \ \ \ \ \ \ \ \mathbf{\Phi }_{p}^{l}=\mathrm{diag}\left( e^{j\theta _{p ,1}^{l}},e^{j\theta _{p ,2}^{l}},\cdots ,e^{j\theta _{p ,M}^{l}} \right) \in \mathbb{C} ^{M\times M}, \theta _{p ,m}^{l}\in \left[ 0,2\pi \right) ,m\in \mathcal{M} ,l\in \mathcal{L} ,p \in \left\{ 0,1 \right\},\tag{1}\\
\label{eq2}
&\ \ \ \ \ \ \ \ \ \ \ \ \ \ \ \ \ \ \ \ \ \ \ \mathbf{\Psi} _{p}^{k}=\mathrm{diag}\left( e^{j\xi _{p ,1}^{k}},e^{j\xi _{p ,2}^{k}},\cdots ,e^{j\xi _{p ,N}^{k}} \right) \in \mathbb{C} ^{N\times N},\xi _{p ,n}^{k}\in \left[ 0,2\pi \right) ,n\in \mathcal{N} ,k\in \mathcal{K} ,p \in \left\{ 0,1 \right\} .\tag{2}\\
\label{eq5}
&\ \ \ \ \ \ \ \ \ \ \ \ \ \ \ \ \ \ \ \ \ \ \ \ \mathbf{T}=\left[ \begin{matrix}
	\mathbf{\Phi }_{0}^{L}&		\mathbf{0}\\
	\mathbf{0}&		\mathbf{\Phi }_{1}^{L}\\
\end{matrix} \right] \left[ \begin{matrix}
	\mathbf{V}^L&		\mathbf{0}\\
	\mathbf{0}&		\mathbf{V}^L\\
\end{matrix} \right] \cdots \left[ \begin{matrix}
	\mathbf{\Phi }_{0}^{2}&		\mathbf{0}\\
	\mathbf{0}&		\mathbf{\Phi }_{1}^{2}\\
\end{matrix} \right] \left[ \begin{matrix}
	\mathbf{V}^2&		\mathbf{0}\\
	\mathbf{0}&		\mathbf{V}^2\\
\end{matrix} \right] \left[ \begin{matrix}
	\mathbf{\Phi }_{0}^{1}&		\mathbf{0}\\
	\mathbf{0}&		\mathbf{\Phi }_{1}^{1}\\
\end{matrix} \right] \left[ \begin{matrix}
	\mathbf{V}^1&		\mathbf{0}\\
	\mathbf{0}&		\mathbf{V}^1\\
\end{matrix} \right] \in \mathbb{C} ^{2M\times 2S},\tag{5}
\\
\label{eq6}
&\ \ \ \ \ \ \ \ \ \ \ \ \ \ \ \ \ \ \ \ \ \ \ \ \mathbf{R}=\left[ \begin{matrix}
	\mathbf{U}^1&		\mathbf{0}\\
	\mathbf{0}&		\mathbf{U}^1\\
\end{matrix} \right] \left[ \begin{matrix}
	\mathbf{\Psi }_{0}^{1}&		\mathbf{0}\\
	\mathbf{0}&		\mathbf{\Psi }_{1}^{1}\\
\end{matrix} \right] \left[ \begin{matrix}
	\mathbf{U}^2&		\mathbf{0}\\
	\mathbf{0}&		\mathbf{U}^2\\
\end{matrix} \right] \left[ \begin{matrix}
	\mathbf{\Psi }_{0}^{2}&		\mathbf{0}\\
	\mathbf{0}&		\mathbf{\Psi }_{1}^{2}\\
\end{matrix} \right] \cdots \left[ \begin{matrix}
	\mathbf{U}^K&		\mathbf{0}\\
	\mathbf{0}&		\mathbf{U}^K\\
\end{matrix} \right] \left[ \begin{matrix}
	\mathbf{\Psi }_{0}^{K}&		\mathbf{0}\\
	\mathbf{0}&		\mathbf{\Psi }_{1}^{K}\\
\end{matrix} \right] \in \mathbb{C} ^{2S\times 2N}.\tag{6}
\end{align*}
\end{small}
\end{figure*}


We consider a DPSIM-assisted HMIMO system, as shown in Fig. \ref{Fig1}, where TX-DPSIM and RX-DPSIM are integrated in the transmitter and receiver, respectively. To facilitate understanding the model, Fig.~\ref{Fig2} shows a schematic diagram of the HMIMO system in the two orthogonal polarization directions. We assume that each layer in DPSIM is an identical square structure, where each layer of TX-DPSIM contains $M$ DPEM units, and each layer of RX-DPSIM contains $N$ DPEM units, satisfying $M\geqslant S$ and $N\geqslant S$. Their respective sets are represented by $\mathcal{M} =\left\{ 1,2,\cdots,M \right\} $ and $\mathcal{N} =\left\{ 1,2,\cdots,N \right\}$. We assume that $2S$ and $\mathcal{S} =\left\{ 1,2,\cdots ,2S \right\} $ denote the number of data streams and the corresponding set, respectively. $L$ and $K$ denote the number of metasurface layers at the transmitter and receiver, while their respective sets are represented by $\mathcal{L} =\left\{ 1,2,\cdots ,L \right\} $  and $\mathcal{K} =\left\{ 1,2,\cdots ,K \right\} $. $p \in \mathcal{P}=\left\{ 0,1 \right\}$ denotes two orthogonal polarization directions. Due to the presence of dual-polarization defects (polarization conversion), we use four matrices (00, 11, 10, and 01) to characterize the channel. These matrices respectively represent the channel from polarization 0 to 0, from polarization 1 to 1, from polarization 0 to 1, and from polarization 1 to 0~\cite{9497725}. Considering that each layer of the metasurface in DPSIM can independently control the phase for different polarization directions, therefore DPSIM can be equivalently treated as two identical and isolated single-polarization SIMs.

The transmission coefficient of the $l$-th layer for TX-DPSIM and the $k$-th layer for RX-DPSIM are defined as~\eqref{eq1} and~\eqref{eq2}, respectively. According to the Rayleigh-Sommerfeld diffraction theory~\cite{Lin2018Alloptical}, the transmission coefficient from the $\tilde{m}$-th DPEM unit on the $(l-1)$-st transmit metasurface layer to the $m$-th DPEM unit on the $l$-th transmit metasurface layer is expressed by
\begin{align*}
\label{3}
&\left[ \mathbf{V}^{l} \right] _{m,\tilde{m}}=\frac{A_t\cos \chi _{m,\tilde{m}}^{l}}{r_{m,\tilde{m}}^{l}}\left( \frac{1}{2\pi r_{m,\tilde{m}}^{l}}-\frac{j}{\lambda} \right) e^{j2\pi r_{m,\tilde{m}}^{l}/\lambda}.\tag{3}
\end{align*}
Similarly, the transmission coefficient from the $n$-th DPEM unit on the $k$-th receive metasurface layer to the $\tilde{n}$-th DPEM unit on the $(k-1)$-st receive metasurface layer is expressed by
\begin{align*}
\label{4}
&\left[ \mathbf{U}^{k} \right] _{\tilde{n},n}=\frac{A_r\cos \zeta _{\tilde{n},n}^{k}}{r_{\tilde{n},n}^{k}}\left( \frac{1}{2\pi r_{\tilde{n},n}^{k}}-\frac{j}{\lambda} \right) e^{j2\pi r_{\tilde{n},n}^{k}/\lambda}. \tag{4}
\end{align*}
Where $r_{m,\tilde{m}}^l$ and $r_{\tilde{n},n}^l$ denote the corresponding transmission distance, $A_t$ and $A_r$ denote the area of each DPEM unit, $\chi_{m,\tilde{m}}^l$ and $\zeta _{\tilde{n},n}^{k}$ represent the angle between the propagation direction and the normal direction of the metasurface layer. Considering that the interaction between electromagnetic waves and environmental scatterers is the main mechanism for changing their polarization state~\cite{5979177}, we assume that the polarization state of the electromagnetic waves remains unchanged during their propagation between DPSIM layers. Therefore, the overall coefficient matrices for TX-DPSIM and RX-DPSIM are given by~\eqref{eq5} and~\eqref{eq6}, respectively.

In summary, the signal reception model for the HMIMO system is given by
\begin{align*}
\label{eq7}
\mathbf{y}=\mathbf{RGTpx}+\mathbf{n},\tag{7}
\end{align*}
where $\mathbf{n}\in \mathbb{C} ^{2S\times 1}$ is the receiver noise vector with distribution $ \mathcal{C} \mathcal{N} \left( 0,\sigma ^2\mathbf{I}_{2S} \right)$, and $\mathbf{x}\in \mathbb{C} ^{2S\times 1}$ is the signal vector satisfying $\mathbb{E} \left\{ \mathbf{xx}^{\mathrm{H}} \right\} =\mathbf{I}_{2S}$. $\mathbf{p}\in \mathbb{C} ^{2S\times 1}$ denotes the vector of transmit powers for different data streams. Considering the spatial correlation among the metasurface DPEM units, We model the channel $\mathbf{G}\in \mathbb{C} ^{2N\times 2M}$ between TX-DPSIM and RX-DPSIM as \cite{9497725}, ~\cite{9714406}
\begin{align*}
\label{eq8}
&\mathbf{G}=\left[ \begin{matrix}
	\mathbf{R}_{\mathrm{RX}}^{1/2}&		\mathbf{R}_{\mathrm{RX}}^{1/2}\\
	\mathbf{R}_{\mathrm{RX}}^{1/2}&		\mathbf{R}_{\mathrm{RX}}^{1/2}\\
\end{matrix} \right] \left[ \begin{matrix}
	\tilde{\mathbf{G}}_{00}&		\tilde{\mathbf{G}}_{01}\\
	\tilde{\mathbf{G}}_{10}&		\tilde{\mathbf{G}}_{11}\\
\end{matrix} \right] \left[ \begin{matrix}
	\mathbf{R}_{\mathrm{TX}}^{1/2}&		\mathbf{R}_{\mathrm{TX}}^{1/2}\\
	\mathbf{R}_{\mathrm{TX}}^{1/2}&		\mathbf{R}_{\mathrm{TX}}^{1/2}\\
\end{matrix} \right],\tag{8}\\
\label{eq9}
&\tilde{\mathbf{G}}_{00},\tilde{\mathbf{G}}_{11}\sim \mathcal{C} \mathcal{N} \left( 0,\left( 1-\epsilon \right)\text{PL}(d)\mathbf{I}_N\otimes \mathbf{I}_M \right), \tag{9}\\
\label{eq10}
&\tilde{\mathbf{G}}_{10},\tilde{\mathbf{G}}_{01}\sim \mathcal{C} \mathcal{N} \left( 0,\epsilon\text{PL}(d)\mathbf{I}_N\otimes \mathbf{I}_M \right),\tag{10}
\end{align*}
where $\tilde{\mathbf{G}}_{qp}\in \mathbb{C} ^{N\times M}$ denotes the independent and identically distributed Rayleigh fading channel from polarization $p$ to polarization $q$. $0\leqslant\epsilon \leqslant 1$ is the proportion of radiated power converted between polarization 1 and polarization 0. $\text{PL}(d)$ is the path loss between the transmitter and receiver. $\mathbf{R}_{\mathrm{TX}} \in \mathbb{C}^{M \times M}$ and $\mathbf{R}_{\mathrm{RX}} \in \mathbb{C}^{N \times N}$ represent the spatial correlation matrix at the TX-DPSIM and that at the RX-DPSIM, respectively. $\mathbf{R}_{\mathrm{TX/RX}}^{1/2}$ represents the square root of the matrix. By considering far-field propagation in an isotropic scattering environment, the spatial correlation matrix can be expressed by~\cite{9110848}
\begin{align*}
\label{eq11}
[\mathbf{R}_{\mathrm{TX}}]_{m,\tilde{m}} &= \text{sinc}(2r_{m,\tilde{m}}/\lambda), \quad \tilde{m} \in \mathcal{M}, m \in \mathcal{M},\tag{11} \\
\label{eq12}
[\mathbf{R}_{\mathrm{RX}}]_{\tilde{n},n} &= \text{sinc}(2r_{\tilde{n},n}/\lambda), \quad n \in \mathcal{N}, \tilde{n} \in \mathcal{N},\tag{12}
\end{align*}
respectively. $r_{m,\tilde{m}}$ and $r_{\tilde{n},n}$ represent the distance between different corresponding DPEM units on a single-layer metasurface.

\section{Problem Formulation and Solution}

\begin{figure*}[b]
\begin{small}
\vspace{-1cm}
\begin{align*}
&\overline{\ \ \ \ \ \ \ \ \ \ \ \ \ \ \ \ \ \ \ \ \ \ \ \ \ \ \ \ \ \ \ \ \ \ \ \ \ \ \ \ \ \ \ \ \ \ \ \ \ \ \ \ \ \ \ \ \ \ \ \ \ \ \ \ \ \ \ \ \ \ \ \ \ \ \ \ \ \ \ \ \ \ \ \ \ \ \ \ \ \ \ \ \ \ \ \ \ \ \ \ \ \ \ \ \ \ \ \ \ \ \ \ \ \ \ \ \ \ \ \ \ \ \ \ \ \ \ \ \ \ \ \ \ \ \ \ \ \ \ \ \ \ \ \ \ \ \ \ \ \ \ \ \ \ \ \ \ \ \ \ \ \ \ \ }\\
\label{eq16}
&\ \ \ \ \ \ \ \ \ x_{\acute{m},s,\tilde{s}}^{l}=\mathbf{R}_{s,:}\mathbf{G}\left[ \begin{matrix}
	\mathbf{\Phi }_{0}^{L}&		\mathbf{0}\\
	\mathbf{0}&		\mathbf{\Phi }_{1}^{L}\\
\end{matrix} \right] \left[ \begin{matrix}
	\mathbf{V}^L&		\mathbf{0}\\
	\mathbf{0}&		\mathbf{V}^L\\
\end{matrix} \right] \dots \left[ \begin{matrix}
	\mathbf{V}^{l+1}&		\mathbf{0}\\
	\mathbf{0}&		\mathbf{V}^{l+1}\\
\end{matrix} \right] _{:,\acute{m}}\left[ \begin{matrix}
	\mathbf{V}^l&		\mathbf{0}\\
	\mathbf{0}&		\mathbf{V}^l\\
\end{matrix} \right] _{\acute{m},:}\dots \left[ \begin{matrix}
	\mathbf{\Phi }_{0}^{1}&		\mathbf{0}\\
	\mathbf{0}&		\mathbf{\Phi }_{1}^{1}\\
\end{matrix} \right] \left[ \begin{matrix}
	\mathbf{V}^1&		\mathbf{0}\\
	\mathbf{0}&		\mathbf{V}^1\\
\end{matrix} \right] _{:,\tilde{s}},\tag{16}\\
\label{eq17}
&\ \ \ \ \ \ \ \ \ y_{\acute{n},s,\tilde{s}}^{k}=\left[ \begin{matrix}
	\mathbf{U}^1&		\mathbf{0}\\
	\mathbf{0}&		\mathbf{U}^1\\
\end{matrix} \right] _{s,:}\left[ \begin{matrix}
	\mathbf{\Psi }_{0}^{1}&		\mathbf{0}\\
	\mathbf{0}&		\mathbf{\Psi }_{1}^{1}\\
\end{matrix} \right] \dots \left[ \begin{matrix}
	\mathbf{U}^k&		\mathbf{0}\\
	\mathbf{0}&		\mathbf{U}^k\\
\end{matrix} \right] _{:,\acute{n}}\left[ \begin{matrix}
	\mathbf{U}^{k+1}&		\mathbf{0}\\
	\mathbf{0}&		\mathbf{U}^{k+1}\\
\end{matrix} \right] _{\acute{n},:}\dots \left[ \begin{matrix}
	\mathbf{U}^K&		\mathbf{0}\\
	\mathbf{0}&		\mathbf{U}^K\\
\end{matrix} \right] \left[ \begin{matrix}
	\mathbf{\Psi }_{0}^{K}&		\mathbf{0}\\
	\mathbf{0}&		\mathbf{\Psi }_{1}^{K}\\
\end{matrix} \right] \mathbf{GT}_{:,\tilde{s}}.\tag{17} 
\end{align*}
\end{small}
\end{figure*}

\begin{algorithm}[!t]
\begin{footnotesize}
    \caption{LGD-WF Algorithm.}
    \KwIn{$E^{\max} $; $P_t$; $\sigma ^2$; $\varsigma$; $\eta$; $\beta$; $\mathbf{V}^l, l \in \mathcal{L}$; $\mathbf{G}$; $\mathbf{U}^k, k \in \mathcal{K}$.}
    \KwOut{$\mathbf{p}$; $\alpha$; $\theta _{p ,m}^{l}$, $m\in \mathcal{M}$, $l\in \mathcal{L}$, $p \in \mathcal{P}$; $\xi _{p ,n}^{k}$, $n\in \mathcal{N}$, $k\in \mathcal{K}$, $p \in \mathcal{P}$.}
    Initialization:\ Generate $\varsigma$ random sets of initial phases $\theta _{p ,m}^{l}$, $\xi _{p ,n}^{k}$, and select the set minimizing $\varGamma$.\\
    \For{$e=1\sim E^{\max} $}{
        \For{$l=1\sim L$}{
            Calculating ${\partial \varGamma} / {\partial \theta _{p ,m}^{l}}$, $m\in \mathcal{M}$, $p \in \mathcal{P}$ by ~\eqref{eq14}; \\
            Normalizing ${\partial \varGamma} / {\partial \theta _{p ,m}^{l}}$, $m\in \mathcal{M}$, $p \in \mathcal{P}$ by ~\eqref{eq18}; \\
            Updating $\theta _{p ,m}^{l}$, $m\in \mathcal{M}$, $p \in \mathcal{P}$ by ~\eqref{eq20};\\
            Updating $\alpha$ by applying ~\eqref{eq22};\\
        }
        \For{$k=1\sim K$}{
            Calculating ${\partial \varGamma} / {\partial \xi _{p ,n}^{k}}$, $n\in \mathcal{N}$, $p \in \mathcal{P}$ by ~\eqref{eq15}; \\
            Normalizing ${\partial \varGamma} / {\partial \xi _{p ,n}^{k}}$, $n\in \mathcal{N}$, $p \in \mathcal{P}$ by ~\eqref{eq19}; \\
            Updating $\xi _{p ,n}^{k}$, $n\in \mathcal{N}$, $p \in \mathcal{P}$ by ~\eqref{eq21};\\
            Updating $\alpha$ by applying ~\eqref{eq22};\\
        }
        Diminishing the learning rate $\eta$ by ~\eqref{eq23};\\
    }
    Power allocation by Water-Filling Algorithm
\end{footnotesize}
\end{algorithm}

For a given channel matrix $\mathbf{G}$, we employ a truncated singular value decomposition (SVD) strategy to realize HMIMO transmission~\cite{9614196}. Specifically, we aim to minimize the difference between the end-to-end channel $\mathbf{H}=\mathbf{RGT}$ and the desired matrix $\mathbf{\Lambda }_{1:2S,1:2S}$, where $\mathbf{\Lambda} =\mathrm{diag}\mathrm{(}\lambda _1,\lambda _2,\cdots ,\lambda _o)$ is the diagonal matrix comprising singular values of $\mathbf{G}$, with $o=\min \left( 2M,2N \right)$ and the singular values ordered as $\lambda _1\geqslant \lambda _2\geqslant \cdots \geqslant \lambda _o$. The problem is formulated as
\begin{align*}
\label{eq13}
\mathcal{P} 1:\,\underset{\theta _{p ,m}^{l},\xi _{p ,n}^{k},\alpha}{\min}\,&\varGamma =\parallel \alpha \mathbf{RGT}-\mathbf{\Lambda }_{1:2S,1:2S}\parallel _{\mathrm{F}}^{2},\tag{13}\\
\mathrm{s}.t.\ \ \  &\left( 1-12 \right) ,\alpha \in \mathbb{C},
\end{align*}
where $\alpha$ is a compensation scaling factor introduced to account for the EE gains of the HMIMO system due to its avoidance of traditional precoding, facilitating a fair comparison of the signal processing performance of SIM and DPSIM~\cite{10158690},~\cite{li2025stackedintelligentmetasurfacesenhancedmimo}.

Inspired by the gradient descent algorithm in \cite{10158690}, we propose the LGD-WF Algorithm to realize HMIMO in the orthogonal dual-polarized domain. The algorithm consists of two parts: the first part solves $\mathcal{P}1$ through layer-by-layer gradient descent, and the other part realizes power allocation for different data streams through the water-filling algorithm~\cite{Shannon1948A}. The core steps are as follows:

\textbf{Step1. Calculate the partial derivatives.} The partial derivatives of $\varGamma$ with respect to $\theta _{p,m}^{l}$ and $\xi _{p,n}^{k}$ are respectively given by
\begin{align*}
\label{eq14}
\frac{\partial \varGamma}{\partial \theta _{p,m}^{l}}=2\sum_{s=1}^{2S}{\sum_{\tilde{s}=1}^{2S}{\Im}}\left[ \left( \alpha e^{j\theta _{p,m}^{l}}x_{\acute{m},s,\tilde{s}}^{l} \right) ^*\left( \alpha [\mathbf{H}] _{s,\tilde{s}}-[\mathbf{\Lambda }]_{s,\tilde{s}} \right) \right],\tag{14}\\
\label{eq15}
\frac{\partial \varGamma}{\partial \xi _{p,n}^{k}}=2\sum_{s=1}^{2S}{\sum_{\tilde{s}=1}^{2S}{\Im}}\left[ \left( \alpha e^{j\xi _{p,n}^{k}}y_{\acute{n},s,\tilde{s}}^{k} \right) ^*\left( \alpha [\mathbf{H}] _{s,\tilde{s}}-[\mathbf{\Lambda }]_{s,\tilde{s}} \right) \right], \tag{15}
\end{align*}
where $x_{\acute{m},s,\tilde{s}}^l$ and $y_{\acute{n},s,\tilde{s}}^k$ are calculated by \eqref{eq16} and \eqref{eq17}, respectively, with $\acute{m}=p\times M+m$ and $\acute{n}=p\times N+n$. $\Im \left( * \right)$ denotes the imaginary part of $*$.

\textbf{Step2. Normalize the partial derivatives.} To mitigate potential gradient explosion and vanishing issues\cite{9142152}, we normalize the partial derivatives for each iteration as follows
\begin{align*}
\label{eq18}
\frac{\partial \varGamma}{\partial \theta _{p,m}^{l}}&\gets \frac{\pi \partial \varGamma}{\partial \theta _{p,m}^{l}}/\max_{m\in \mathcal{M} ,p\in \mathcal{P}} \left( \left| \frac{\partial \varGamma}{\partial \theta _{p,m}^{l}} \right| \right) ,\tag{18}  \\
\label{eq19}
\frac{\partial \varGamma}{\partial \xi _{p,n}^{k}}&\gets \frac{\pi \partial \varGamma}{\partial \xi _{p,n}^{k}}/\max_{n\in \mathcal{N} ,p\in \mathcal{P}} \left( \left| \frac{\partial \varGamma}{\partial \xi _{p,n}^{k}} \right| \right) .\tag{19} 
\end{align*}

\textbf{Step3. Update the phase shifts.} Our phase update strategies for TX-DPSIM and TX-DPSIM are respectively
\begin{align*}
\label{eq20}
\theta _{p,m}^{l}&\gets \theta _{p,m}^{l}-\eta \frac{\partial \varGamma}{\partial \theta _{p,m}^{l}},\tag{20} \\
\label{eq21}
\xi _{p,n}^{k}&\gets \xi _{p,n}^{k}-\eta \frac{\partial \varGamma}{\partial \xi _{p,n}^{k}},\tag{21}
\end{align*}
where $\eta >0$ denotes the learning rate that determines the step size at each iteration.

\textbf{Step4. Update the scaling factor.} Given a fixed set of $\theta _{p,m}^{l}$ and $\xi _{p,n}^{k}$, we can obtain the optimal solution for $\alpha$ through the least squares method as
\begin{align*}
\label{eq22}
\alpha = \frac{\mathrm{Tr}\left( \mathbf{T}^{\mathrm{T}}\mathbf{G}^{\mathrm{T}}\mathbf{R}^{\mathrm{T}}\mathbf{\Lambda }_{1:2S,1:2S} \right)}{\parallel \mathbf{RGT}\parallel _{\mathrm{F}}^{2}}.\tag{22}
\end{align*}

\textbf{Step5. Update the learning rate.} To alleviate issues caused by overshooting during gradient descent, we introduce a negative exponential decay strategy to reduce the learning rate during the phase iterative update process~\cite{LeCun2015Deep}. The strategy is specified as
\begin{equation*}
\label{eq23}
\eta \leftarrow \eta\beta,\tag{23}
\end{equation*}
where $0<\beta<1$ is a hyperparameter controlling the decay.

\textbf{Step6. Power allocation (Water-Filling Algorithm).} To maximize SE, we can obtain the optimal power allocation coefficients by applying the water-filling algorithm. Specifically,
\begin{align*}
\label{eq24}
\mathbf{p}=\left( \tau -\frac{\sigma ^2}{\mathrm{diag}\left( \mathbf{\Lambda }_{1:2S,1:2S}\mathbf{\Lambda }_{1:2S,1:2S} \right)} \right) ^+, \tag{24}
\end{align*}
where $\tau$ is a threshold value satisfying the total transmit power constraint, $\left\| \mathbf{p} \right\| _1=P_t$ with $P_t$ denoting the total available power at the transmitter, which can be obtained by utilizing the bisection method, while $\sigma^2$ represents the average noise power at the receiver.

\section{Simulation Results}

\begin{table}[t]
    \centering
    \caption{system parameters}
    \label{table1}

    \begin{tabular}{ll} 
        \toprule
        HMIMO system parameters parameter & Value \\
        \midrule
        Number of data streams ($2S$) & 6 \\
        Number of layers of TX-DPSIM ($L$) & 3 \\
        Number of layers of RX-DPSIM ($K$) & 3 \\
        Number of DPEM units per layer of TX-DPSIM ($M$) & 100 \\
        Number of DPEM units per layer of RX-DPSIM ($N$) & 100 \\
        Spacing of DPEM units in DPSIM ($r_{\mathrm{DPEM}}$) & $\lambda/2$ \\
        TX-DPSIM thickness ($D_{tx}$) & 0.05 m \\
        RX-DPSIM thickness ($D_{rx}$) & 0.05 m \\
        TX-DPSIM Layer Spacing ($d_{tx}$) & $D_{tx}/L$ \\
        RX-DPSIM Layer Spacing ($d_{rx}$) & $D_{rx}/K$ \\
        Tx-Rx distance ($d$) & 250 m \\
        Transmit power ($P_t$) & 20 dBm \\
        Receiver noise ($\sigma^2$) & $-110$ dBm \\
        Frequency ($f$) & 28 GHz \\
        Wavelength ($\lambda$) & 10.7 mm \\
        Polarization conversion power ratio ($\epsilon$) & 0.2\\
        Path loss reference distance ($d_0$) & 1 m \\
        Path loss exponent ($b$) & 3.5 \\
        Path loss shadowing fading variance ($\delta$) & 9 dB \\
             
        \toprule
        LGD-WF algorithm parameters parameter & Value \\
        \midrule
        Number of randomizations for initialization ($\varsigma$) & 100 \\
        Maximum iterations ($E^{\max}$) & 20 \\
        Initial learning rate ($\eta_0$) & 0.1 \\
        Decay parameter ($\beta$) & 0.5 \\
        Monte Carlo trials & 100 \\
        Initial power scaling factor ($\alpha_0$) & 1 \\
        \bottomrule
        
    \end{tabular}
\end{table}

Since the considered HMIMO system does not perform precoding on the data streams, the residual signal between data streams is directly treated as interference. Therefore, to evaluate the performance of the entire HMIMO system, we define the normalized mean squared error (NMSE) $\Delta $ in \eqref{eq25}, the actual SE $\eta_{\mathrm{SE}}$ in \eqref{eq26}, the theoretical SE upper bound $\eta_{\mathrm{SE}}^{\mathrm{ub}}$in \eqref{eq27}, and the EE $\eta_{\mathrm{EE}},\eta_{\mathrm{EE}}^{\mathrm{ub}}$ in \eqref{eq28}.
\begin{align*}
\label{eq25}
&\Delta = \mathbb{E} \left( \frac{\|\alpha \mathbf{RGT} - \boldsymbol{\Lambda}_{1:2S,1:2S}\|^2_{\mathrm{F}}}{\|\boldsymbol{\Lambda}_{1:2S,1:2S}\|^2_{\mathrm{F}}} \right),\tag{25}\\
\label{eq26}
&\eta _{\mathrm{SE}}=\sum_{s=1}^{2S}{\log _2}\left( 1+\frac{\left[ \mathbf{p} \right] _s|\alpha \left[ \mathbf{H} \right] _{s,s}|^2}{\sum_{\tilde{s}\ne s}^{2S}{\left[ \mathbf{p} \right] _{\tilde{s}}}|\alpha \left[ \mathbf{H} \right] _{s,\tilde{s}}|^2+\sigma ^2} \right),\tag{26}\\
\label{eq27}
&\eta _{\mathrm{SE}}^{_{\mathrm{ub}}}=\sum_{s=1}^{2S}{\log _2}\left( 1+\frac{\left[ \mathbf{p} \right] _s\left[ \mathbf{\Lambda } \right] _{s}^{2}}{\sigma ^2} \right),\tag{27}\\
\label{eq28}
&\eta _{\mathrm{EE}}=\small{\frac{\eta _{\mathrm{SE}}}{P_t}},\ \eta _{\mathrm{EE}}^{\mathrm{ub}}=\small{\frac{\eta _{\mathrm{SE}}^{_{\mathrm{ub}}}}{P_t}}.\tag{28}
\end{align*}

The path loss between the transmitter and the receiver is modeled by~\cite{7109864}
\begin{align*}
\label{eq29}
\text{PL}(d) = \text{PL}(d_0) + 10b \log_{10} \left(\frac{d}{d_0}\right) + X_\delta,\ d \geq d_0,\tag{29}
\end{align*}
where $\text{PL}(d_0) = 20 \log_{10} \left(\frac{4\pi d_0}{\lambda}\right) \text{ dB}$ is the free space path loss at the reference distance $d_0$, $b$ represents the path loss exponent, $X_\delta$ is a zero mean Gaussian random variable with a standard deviation $\delta$, characterizing the large-scale signal fluctuations of shadow fading. Furthermore, the HMIMO model parameters and the LGD-WF algorithm parameters are detailed in Table~\ref {table1}. Unless otherwise specified, all results are based on these parameter settings.

\begin{figure}[!t]
\setlength{\abovecaptionskip}{-0.3cm}
\centering
\includegraphics[width=2.85in]{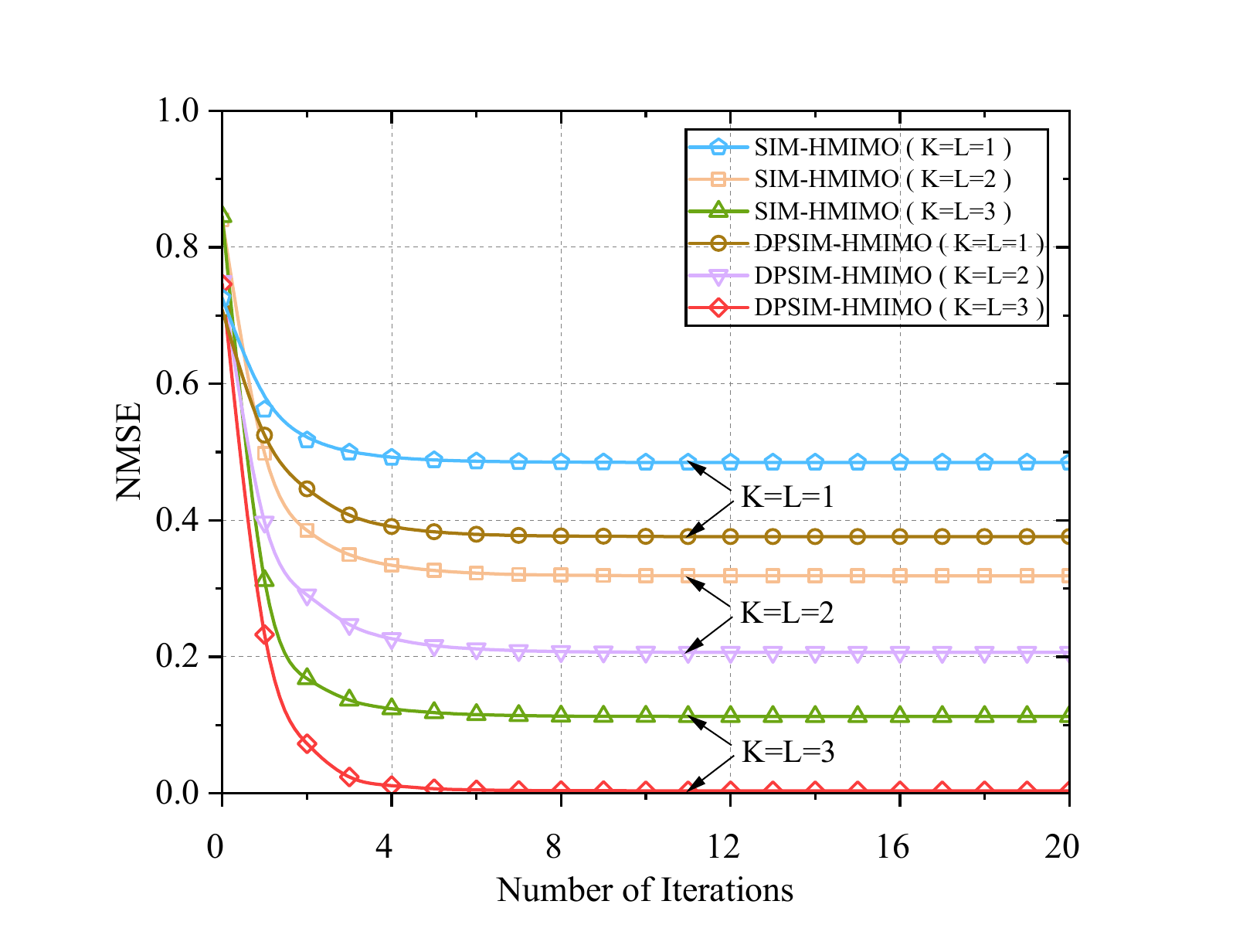}
\caption{NMSE convergence diagram with algorithm iteration.}
\label{Fig3}
\vspace{-0.6cm}
\end{figure}

\begin{figure} [t]
	\centering
        \captionsetup[subfigure]{justification=centering}
	\subfloat[DPSIM, $L=K=1$]{
 		\includegraphics[scale=0.18]{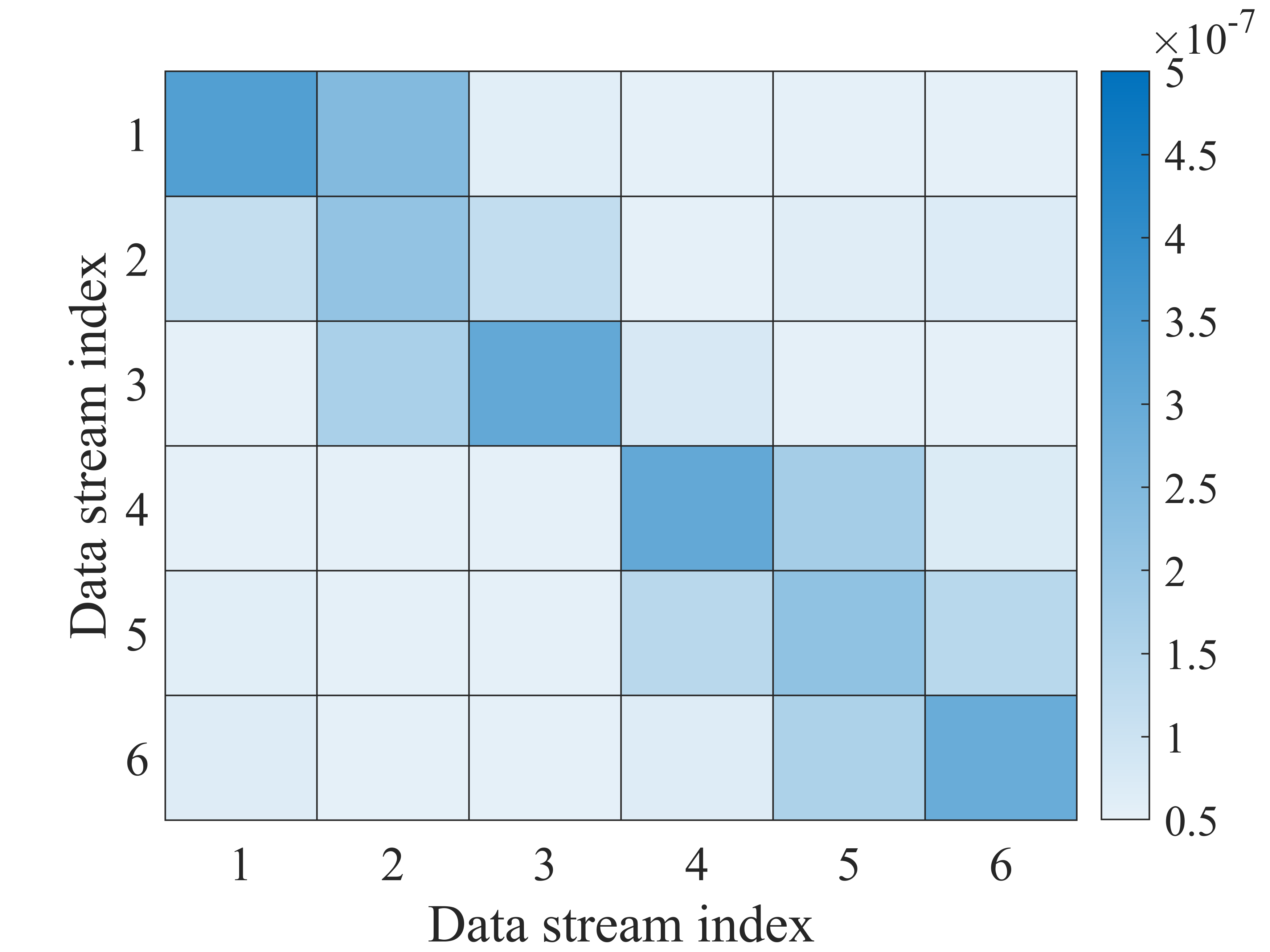}\label{subfig:a}}
	\subfloat[DPSIM, $L=K=2$]{
		\includegraphics[scale=0.18]{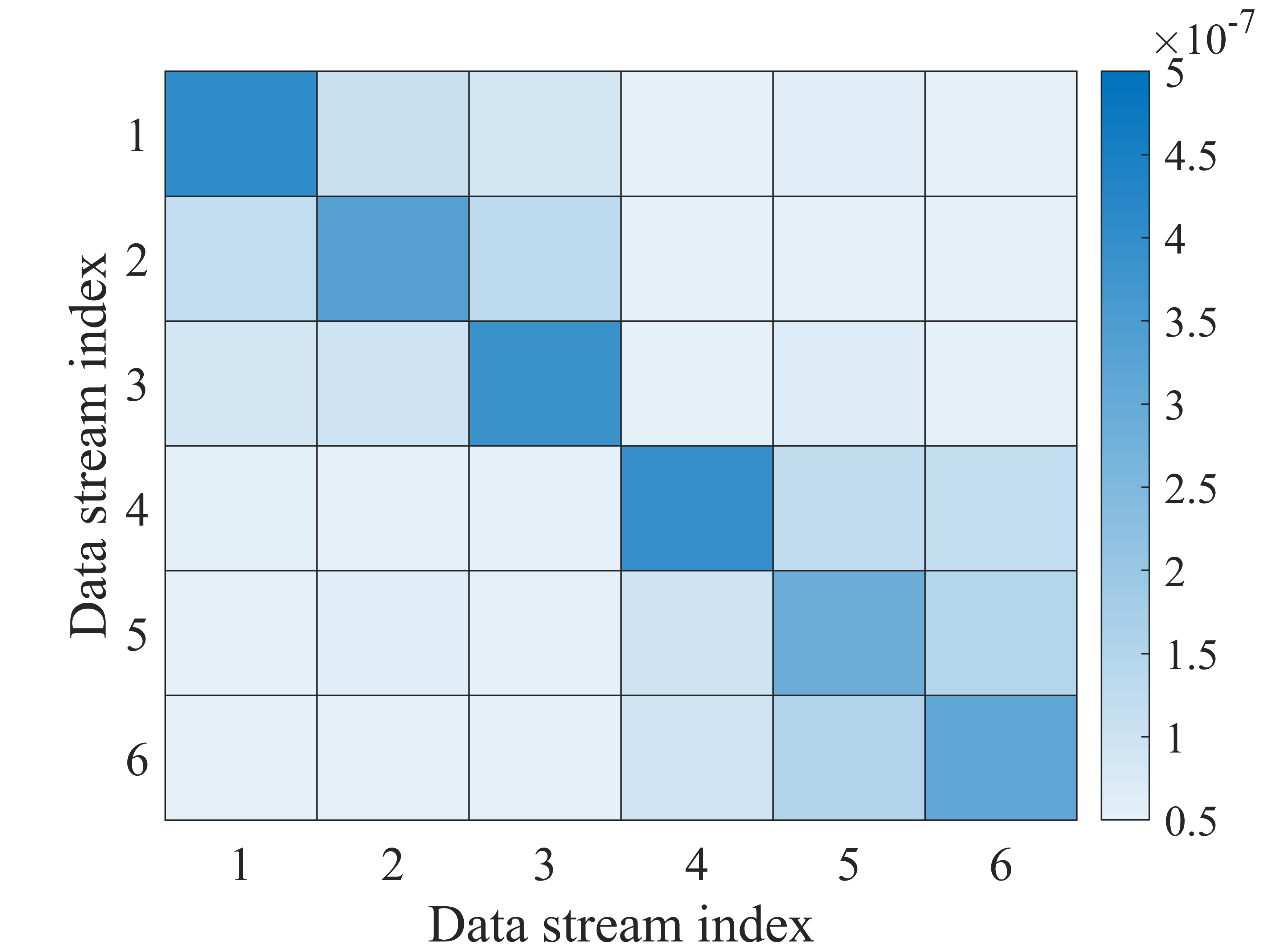}\label{subfig:b}}
        \subfloat[DPSIM, $L=K=3$]{
		\includegraphics[scale=0.18]{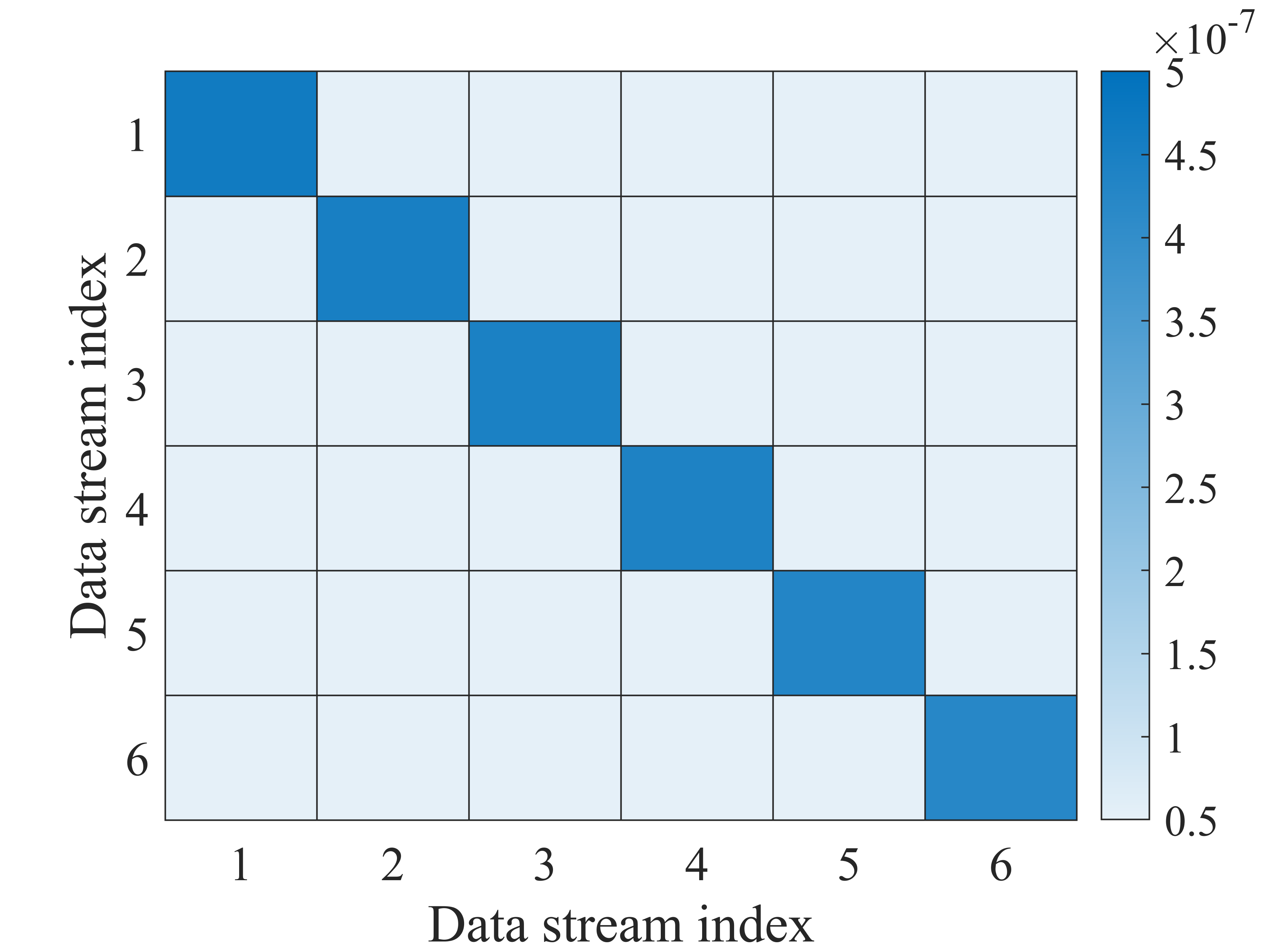}\label{subfig:c}}
        \\
        \vspace{-10pt}
        \subfloat[SIM, $L=K=1$]{
	  \includegraphics[scale=0.18]{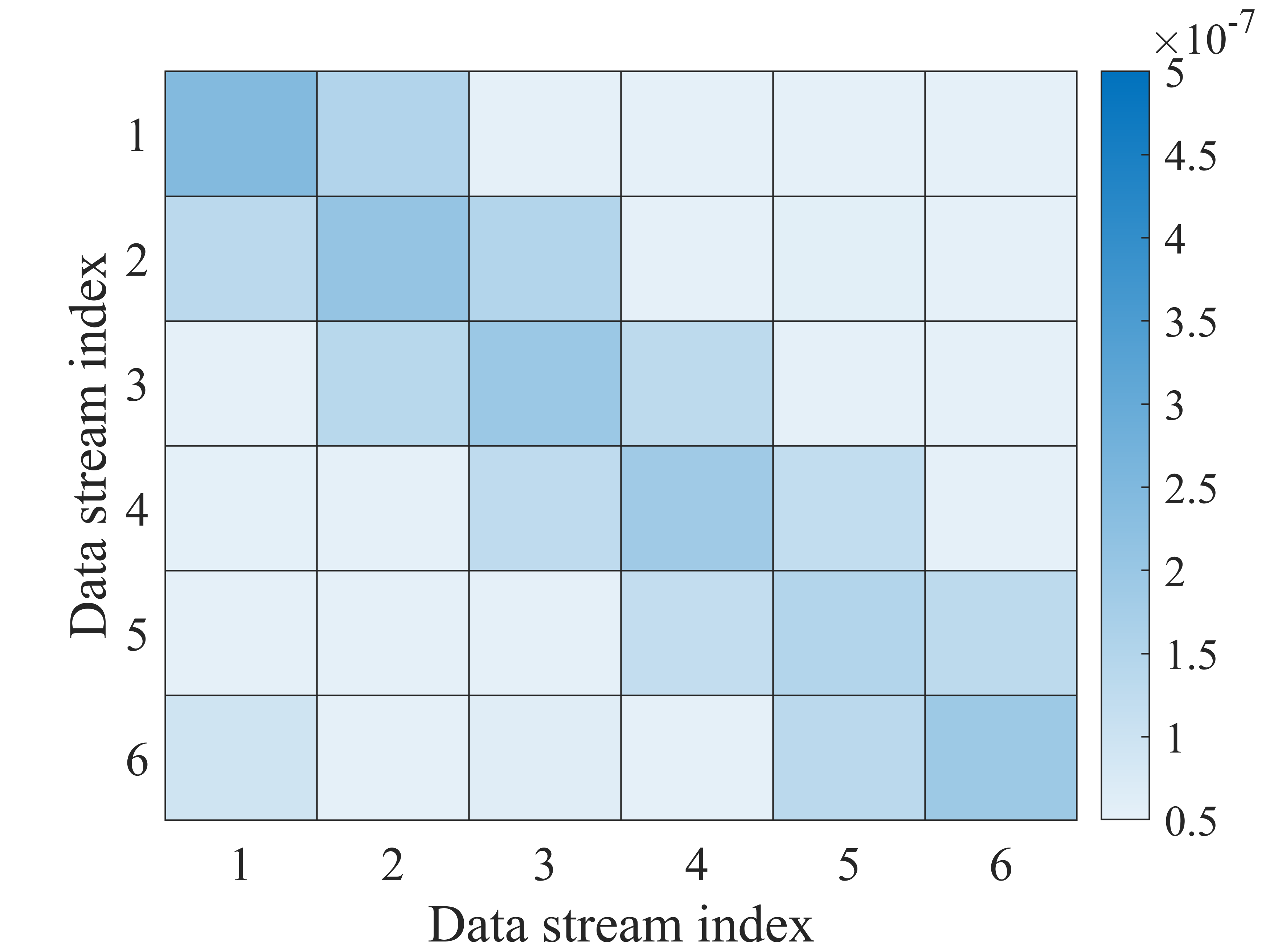}\label{subfig:d}}
        \subfloat[SIM, $L=K=2$]{
	  \includegraphics[scale=0.18]{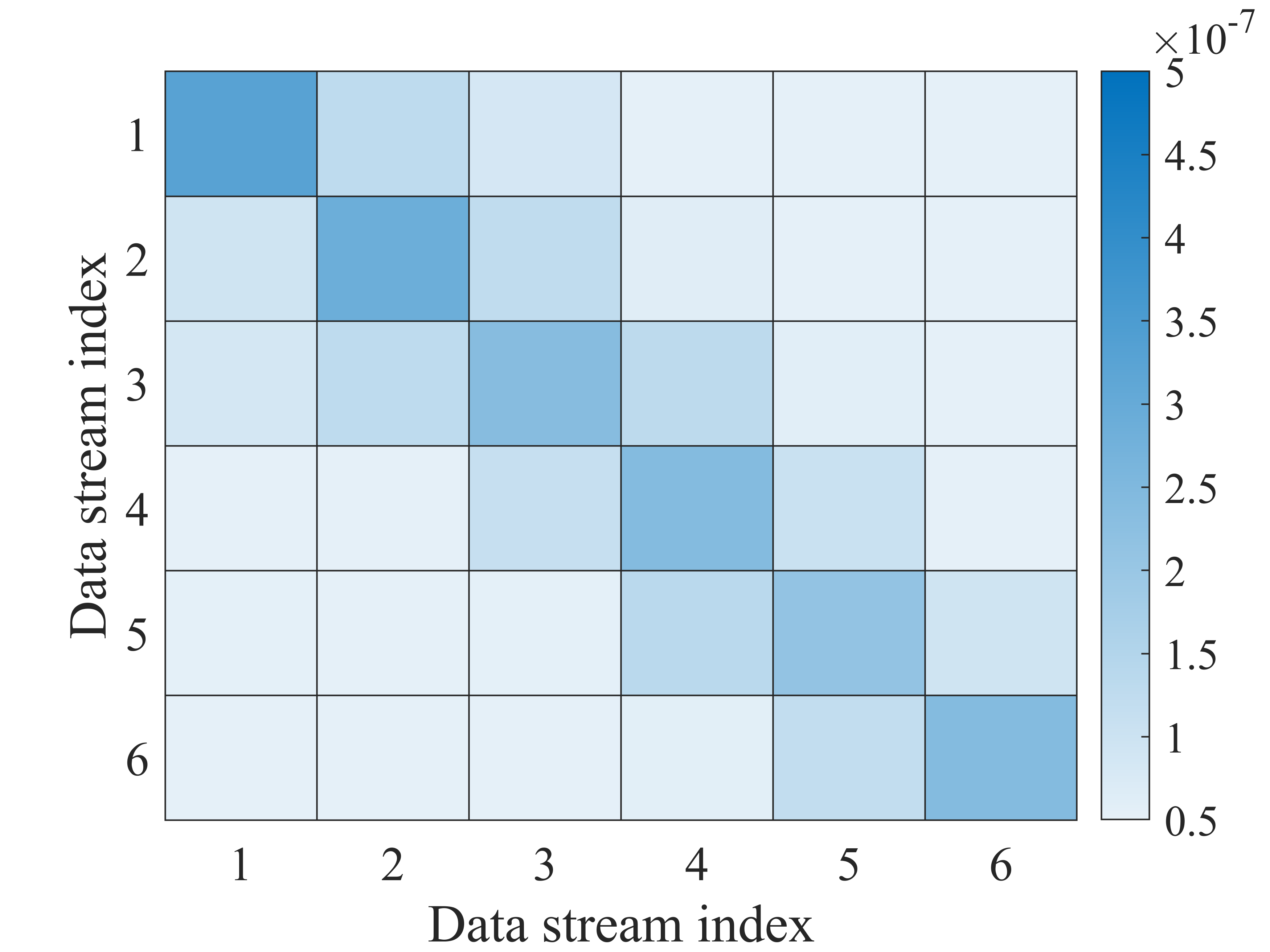}\label{subfig:d}}
        \subfloat[SIM, $L=K=3$]{
	  \includegraphics[scale=0.18]{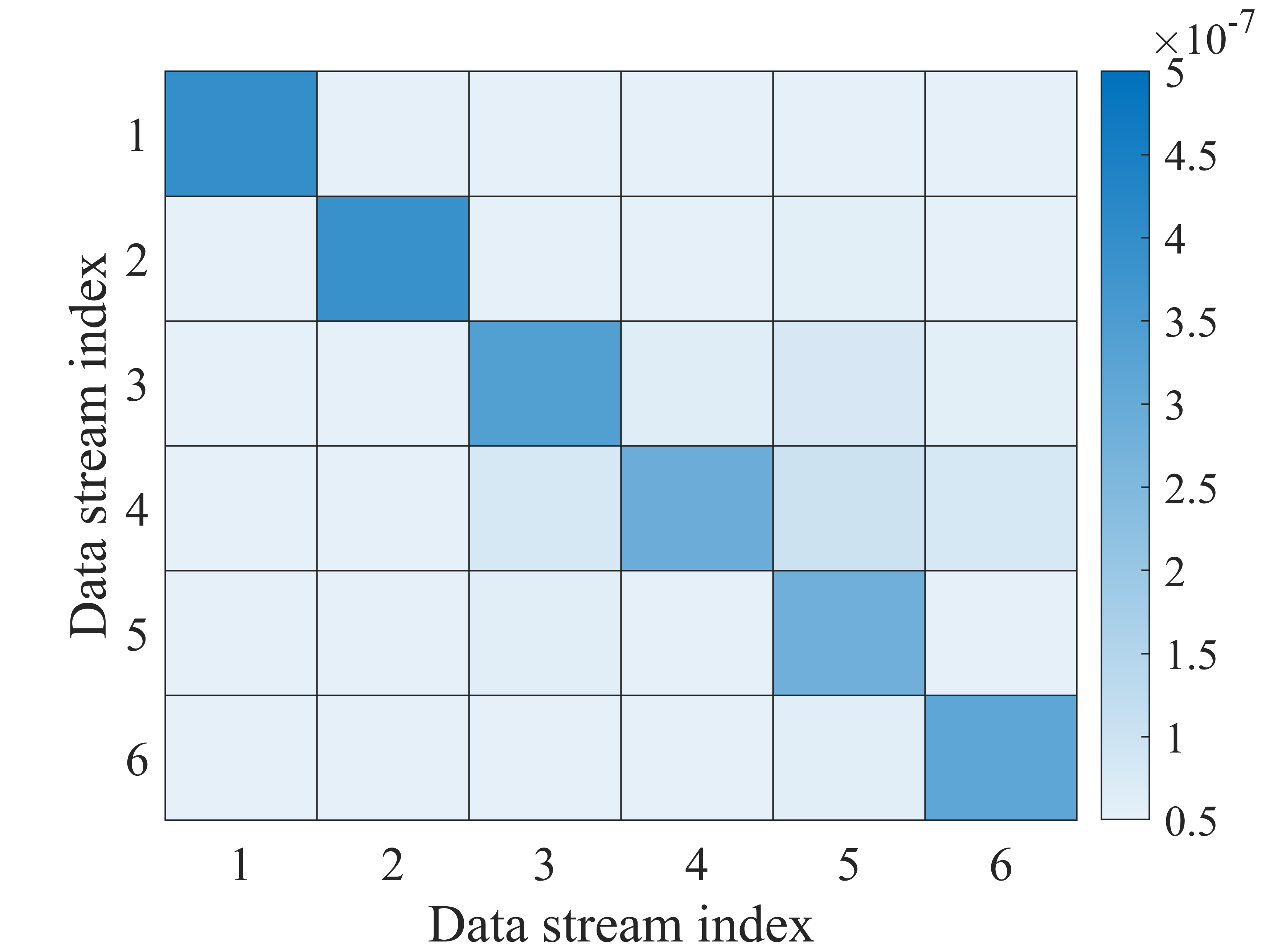}\label{subfig:e}}
    \caption{End-to-end spatial channel matrix.}
\label{Fig4}
\vspace{-0.05cm}
\end{figure}

Fig. \ref{Fig3} and \ref{Fig4} respectively show the NMSE iteration plot and the final optimized end-to-end channel matrix obtained by optimizing SIM/DPSIM using the LGD-WF algorithm. We observe that the NMSE monotonically decreases and converges rapidly with algorithm iterations. With the same number of layers, DPSIM achieves better channel matrix fitting performance compared to SIM. Increasing the number of metasurface layers can enhance the signal processing capability of SIM/DPSIM and better suppress ISI.



\begin{figure}[!t]
\setlength{\abovecaptionskip}{-0.3cm}
\centering
\includegraphics[width=2.85in]{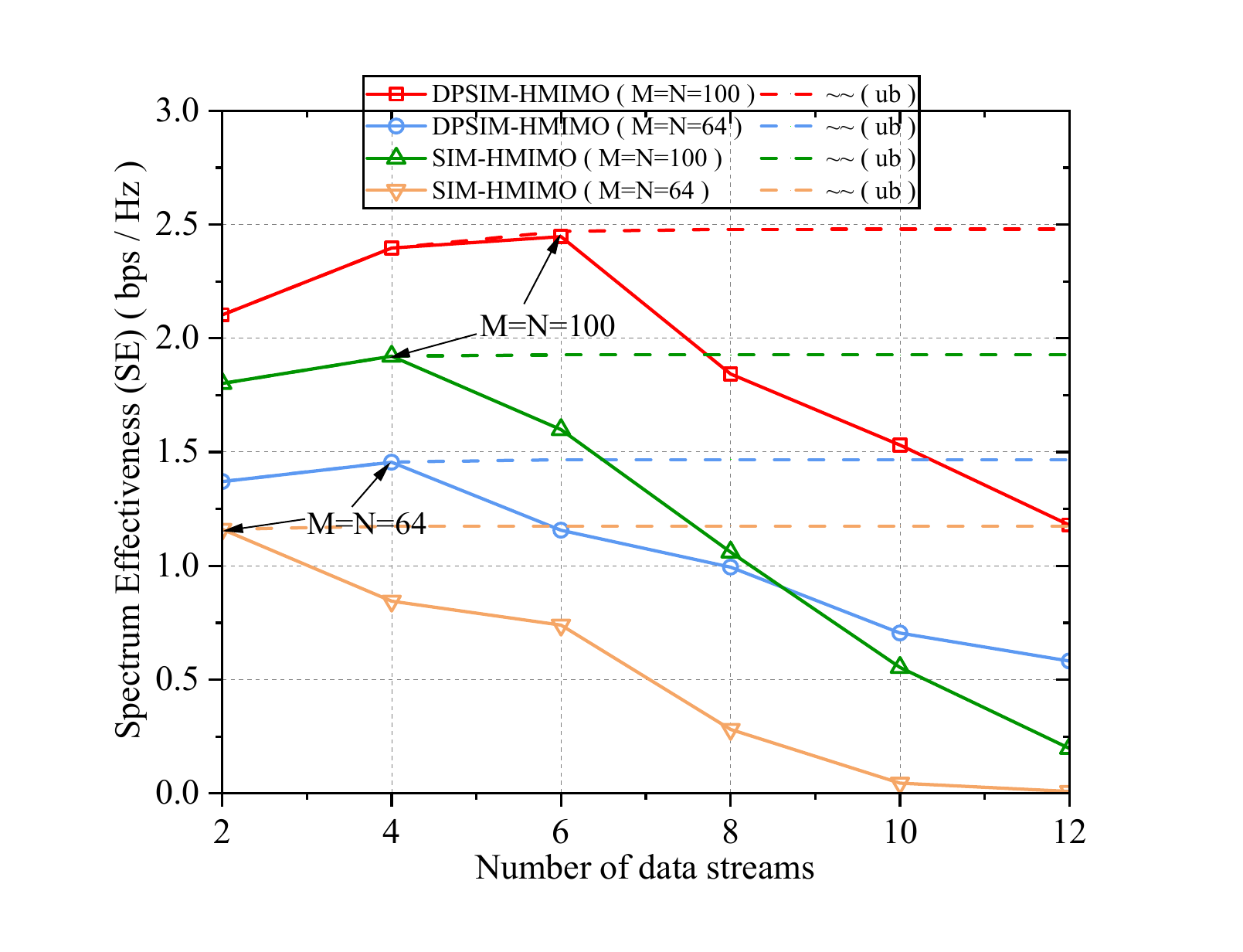}
\caption{The HMIMO system's SE varies with the number of data streams. ($\sim \sim (\mathrm{ub})$ represents the corresponding theoretical upper bound.)}
\label{Fig5}
\vspace{-0.75cm}
\end{figure}

\begin{figure}[!t]
\setlength{\abovecaptionskip}{-0.3cm}
\centering
\includegraphics[width=2.85in]{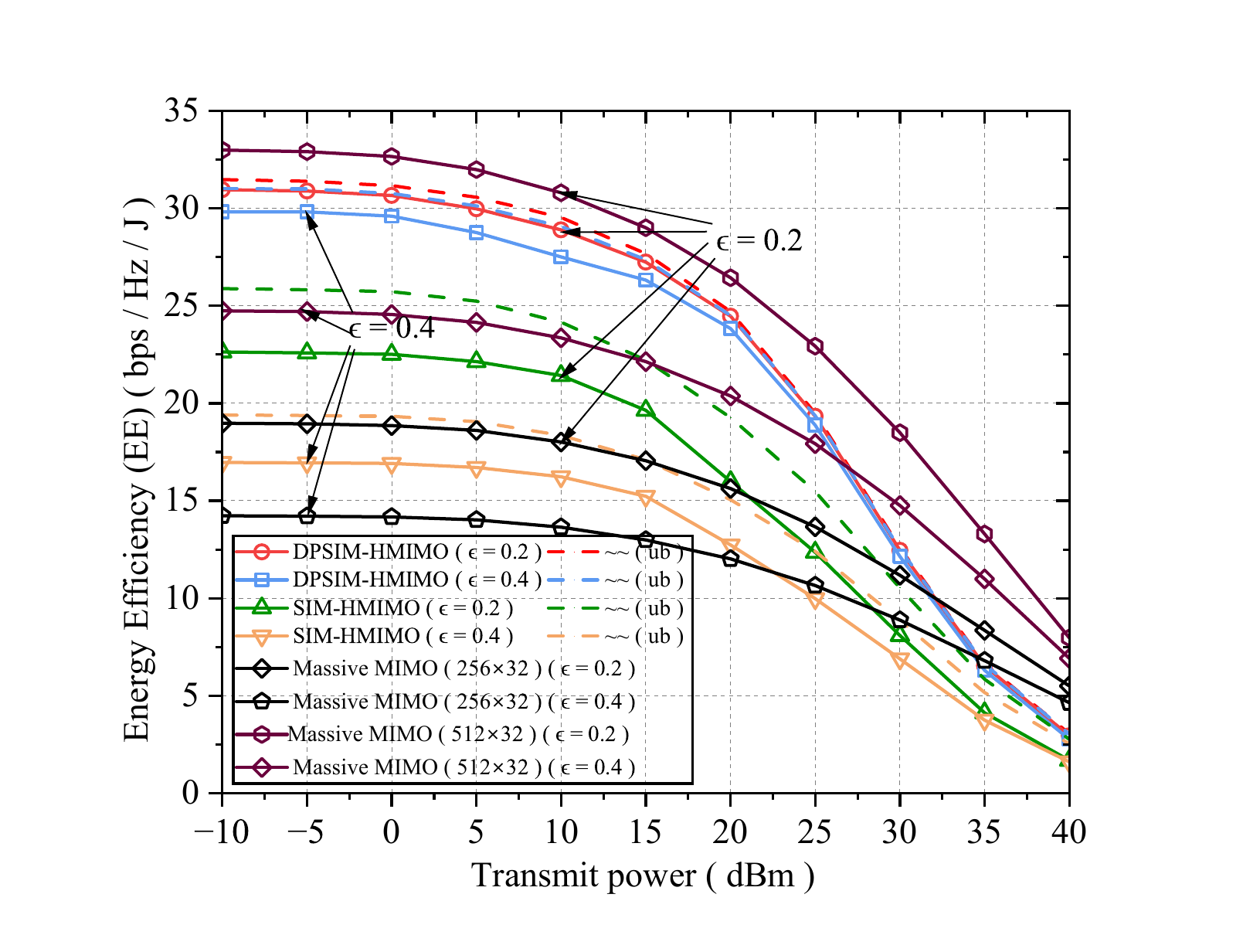}
\caption{The HMIMO system's EE varies with the transmission power change. ($\sim \sim (\mathrm{ub})$ represents the corresponding theoretical upper bound.)}
\label{Fig6}
\end{figure}

Fig. \ref{Fig5} shows the relationship between SE and the number of data streams. We found that appropriately increasing the number of data streams can enhance the SE of HMIMO systems. However, excessive data streams will induce severe ISI due to the limited signal processing capability of SIM/DPSIM, resulting in actual SE far below the theoretical upper bound. The greater the number of metasurface units per layer in SIM/DPSIM, the more interference-free data streams can be supported, resulting in higher SE for the HMIMO system. Compared to SIM-assisted HMIMO, DPSIM-assisted HMIMO can enable interference-free transmission of more data streams.


Fig. \ref{Fig6} shows the variation of EE with total transmit power. We found that increasing transmit power leads to a decrease in EE, and the EE of SIM/DPSIM-aided $6 \times 6$ HMIMO systems is significantly higher than that of $256 \times 32$ traditional massive MIMO systems. Owing to DPSIM providing better ISI suppression, the EE of DPSIM-assisted HMIMO systems is higher than that of SIM-assisted systems. The system EE for a polarization conversion power factor $\epsilon=0.4$ is lower than that for $\epsilon=0.2$. This is because as $\epsilon$ approaches 0.5, the isolation between the two polarization directions of the channel degrades, leading to increased ISI. Furthermore, the DPSIM-assisted system demonstrates comparable EE across different $\epsilon$ channels, approaching its theoretical upper bound in all cases. This indicates that DPSIM maintains robust signal processing capabilities in dual-polarized HMIMO systems.

\section{Conclusion}

To further improve HMIMO system performance within the same spatial constraints, we propose the design of a dual-polarized SIM (DPSIM), and investigate the performance of the DPSIM-assisted HMIMO system. Simulation results demonstrate that, under the same number of metasurface layers and unit size, DPSIM-assisted HMIMO systems suppress ISI more effectively and achieve higher SE and EE than SIM-assisted systems. In the future, we will further explore its potential in multi-user HMIMO systems.

\bibliographystyle{IEEEtran}
\bibliography{references}

\end{document}